# Fourier transform spectrometer on silicon with thermo-optic non-linearity and dispersion correction

Mario C. M. M. Souza[1,2], Andrew Grieco[1], Newton C. Frateschi[2], Yeshaiahu Fainman[1,*]

The integration of miniaturized optical spectrometers into mobile platforms will have unprecedented impact on applications ranging from unmanned aerial vehicles (UAVs) to mobile phones. To address this demand, silicon photonics stands out as a platform capable of delivering compact and cost-effective devices. The Fourier transform spectrometer (FTS) is largely used in free-space spectroscopy, and its implementation in silicon photonics will contribute to bringing broadband operation and fine resolution to the chip scale. The implementation of an integrated silicon photonics FTS (Si-FTS) must nonetheless take into account effects such as waveguide dispersion and non-linearity of refractive index tuning mechanisms. Here we present the modeling and experimental demonstration of a silicon-on-insulator (SOI) Si-FTS with integrated microheaters. We show how the power spectral density (PSD) of a light source and the interferogram measured with the Si-FTS can be related through a simple Fourier transform (FT), provided the optical frequency and time delay are corrected to account for dispersion, thermo-optic non-linearity and thermal expansion. We calibrate the Si-FTS, including the correction parameters, using a tunable laser source and we successfully retrieve the PSD of a broadband source. The aforementioned effects are shown to effectively enhance the Si-FTS resolution when properly accounted for. Finally, we discuss the Si-FTS resilience to chip-scale fabrication variations, a major advantage for large-scale manufacturing. Providing design flexibility and robustness, the Si-FTS is poised to become a fundamental building-block for on-chip spectroscopy.

In recent years, significant efforts have been directed towards the realization of miniaturized optical spectrometers for *in-situ* spectral analysis in numerous areas of science and technology.[1–16] The widespread use of optical spectroscopy from remote sensing[4,7] and planetary sciences[12] to medical research[17] and pharmaceutical processes[18] strongly relies on the large absorption and/or reflection cross-sections of many compounds in the near-infrared (NIR) and mid-infrared (MIR) range.

Silicon-based devices operate in such range and the substantial progress in integrated silicon photonics design and fabrication can therefore be leveraged to develop miniaturized spectrometers for mobile platforms. Typical silicon-on-insulator (SOI) waveguides can operate in the 1.1–4 $\mu$m wavelength range, limited by the silicon band edge at short-wavelengths and by the oxide absorption at long wavelengths. The incorporation of additional CMOS-compatible materials to the mainstream fabrication process, including silicon nitride[4,19] for short wavelengths and germanium-on-silicon[4,20,21] for long wavelengths, promises to significantly extend this window of operation. Moreover, the possibility of monolithic integration of spectrometers and photodetectors is a valuable advantage of photonic integration, promising high signal-to-noise ratio (SNR) and increased sensitivity. Adding heterogeneously integrated light sources[22–24], all the optical components required for a fully functional spectrometer can be realized in a single chip. Finally, the access to multi-project-wafer (MPW) services through silicon photonics foundries provides a cost-effective path to developing robust high-performance devices[4,25–27].

A large variety of integrated photonic spectrometer designs has been recently investigated. These include dispersive devices such as arrayed waveguide gratings (AWG)[4,5] and cavity-enhanced spectrometers[6], spatial heterodyne spectrometers (SHS)[7–10] based on arrays of interferometers, and stationary wave integrated Fourier transform spectrometers (SWIFTS)[11–13]. In the latter, a spatial – rather than temporal – interferogram is formed by the standing wave pattern generated by the interference of two counter-propagating beams inside a waveguide. Designs based on the traditional Fourier transform spectrometer[28], in which a temporal interferogram is measured varying the optical path between the arms of an interferometer, have also been investigated using micro-electro-mechanical systems (MEMS)[14,15] and lithium-niobate planar photonic circuits[16].

Investigations of the traditional FTS design in the silicon photonics platform have been, however, surprisingly scarce[29]. Although one can only speculate about the reasons for such scarcity, some challenges can be identified when considering the requirements for a silicon photonics-based FTS (Si-FTS). First, the optical path difference between the arms of the interferometer is achieved tuning the refractive index rather than changing the physical length, thus an index tuning mechanism capable of large index changes must be used for high spectral resolution. Fortunately, the thermo-optic effect can deliver large index changes of more than $10^{-2}$ for temperature differences around 100 K. For large temperatures, however, thermo-optic non-linearity and thermal expansion of the waveguide become important. Second, silicon waveguides are highly dispersive. As a consequence, the thermo-optic effect will also present strong dispersion. This translates to each optical frequency effectively experiencing a different change in optical path.

In this article, we demonstrate the implementation of a Si-FTS on the SOI platform with integrated microheaters. We show that the issues related to thermo-optic non-linearity, thermal expansion and dispersion can be properly understood, tackled and, moreover, ultimately result in enhanced performance.

## Experimental device

The device consists of a standard Mach-Zehnder Interferometer (MZI) integrated with metal microheaters fabricated in full compatibility with standard silicon photonics foundry processes (Fig.1). The external light is butt-coupled into and out of the chip using inverse tapers and adiabaticaly transitions to the highly confined quasi-TE mode of the access strip waveguide before splitting in the two arms of the interferometer and subsequently recombining into the output waveguide through broadband y-branch couplers (Fig.1e)[30]. The output light is coupled out of chip directly into a photodetector. Each arm of the MZI consists of a spiral (Fig.1d) with total length of 30.407 mm and is covered by independently actuated nichrome microheaters. The device total footprint is 1 mm$^2$.

1. Department of Electrical and Computer Engineering, University of California, San Diego, 9500 Gilman Dr., La Jolla, California 92023, USA.
2. "Gleb Wataghin" Physics Institute, University of Campinas, Campinas, SP 13083-970, Brazil.
*email: fainman@ece.ucsd.edu

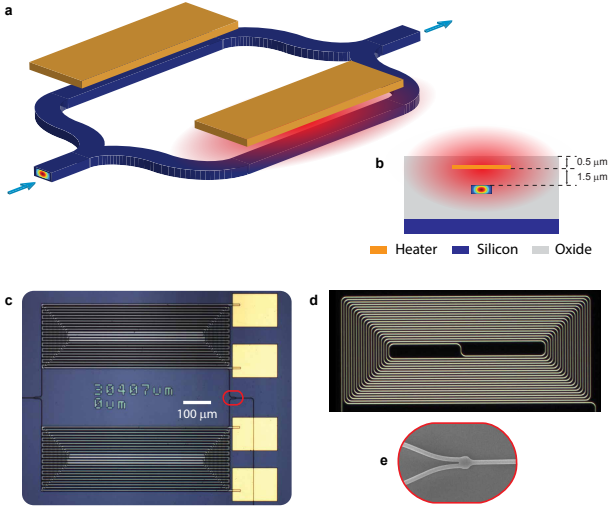

Figure 1: **On-chip Fourier Transform Spectrometer. a.** Schematic of the silicon MZI with integrated microheaters and **b.** device cross-section. **c.** Optical micrography of the experimental device with a total footprint of 1 mm². **d.** Dark field optical micrography of the MZI arm. **e.** SEM image of the broadband power splitter/combiner.

## Spectrometer modeling

In this section, we present the main results that allow to define a modified FT relation between the varying optical power at the output of the MZI, $I$, and the power spectral density of the incoming light, $\text{PSD}(\nu)$. A detailed discussion is provided in Supplementary Section 2.

The operation of the Si-FTS includes a simple data acquisition step consisting of measuring the output power as a function of the phase difference $\Delta\phi$ between the two arms of the MZI. The $\Delta\phi$-dependent term is given by

$$I(\Delta\phi) = \int_{-\infty}^{+\infty} T(\nu)\text{PSD}(\nu)e^{j\Delta\phi(\nu)}d\nu \quad (1)$$

where $\nu$ is the optical frequency and $T(\nu)$ is the transfer function of the MZI – ideally 1. The phase difference is

$$\Delta\phi(\nu) = \frac{2\pi\nu}{c}\left[n_{\text{eff},1}(\nu)L_1 - n_{\text{eff},2}(\nu)L_2\right] \quad (2)$$

where $c$ is the speed of light, $n_{\text{eff},i}$ and $L_i$ are the effective index and the total length of arm $i$.

The discussion is facilitated by first considering the response of an idealized device. In this case $T(\nu) = 1$, the two arms are identical with length $L$, the effective indices are identical and dispersionless, $n_{\text{eff},i}(\nu) \equiv n_{\text{eff}}$, and the effective index change due to temperature change $\Delta T$ depends only on a linear thermo-optic coefficient (TOC) $\partial_T n$, such that $\Delta n_{\text{eff}} = \partial_T n\Delta T$. We use a contracted notation for partial derivatives, $\frac{\partial n_{\text{eff}}}{\partial x} \equiv \partial_x n$. The time delay between the arms of the MZI is defined as $\tau = \frac{L}{c}\partial_T n\Delta T$ and the phase difference is simply

$$\Delta\phi(\nu) = 2\pi\nu\tau. \quad (3)$$

The phase difference in the form $2\pi \times frequency \times delay$ establishes a direct FT relation between $I(\tau)$ and $\text{PSD}(\nu)$, with the conjugate variables $\nu$ and $\tau$,

$$I(\tau) = \int_{-\infty}^{+\infty} \text{PSD}(\nu)e^{j2\pi\nu\tau}d\nu = \mathcal{F}\left[\text{PSD}(\nu)\right] \quad (4)$$

where $\mathcal{F}[\;]$ denotes the Fourier Transform. Thus, $\text{PSD}(\nu)$ can be directly obtained from the inverse FT (IFT) of the interferogram,

$$\text{PSD}(\nu) = \int_{-\infty}^{+\infty} I(\tau)e^{-j2\pi\nu\tau}d\tau = \mathcal{F}^{-1}\left[I(\tau)\right] \quad (5)$$

In practice, the Si-FTS with thermal tuning includes other effects that must be taken into account. First, the strong mode dispersion of silicon waveguides causes significant frequency dependence on the effective index. Second, a large temperature excursion is required to achieve large phase unbalances and the non-linearity of the thermo-optic response must be considered. The large temperature excursion also induces changes in the arm length ($\Delta L$) due to thermal expansion. Finally, chip-scale variability [19,25] and fabrication imperfections often introduce small differences between the two arms of the MZI, identical by design. Such variations may affect the arm length ($\delta L$) as well as the effective index ($\delta n(\nu)$). Assuming the heater on top of arm 1 ($H_1$) is actuated and including the deviations from the designed parameters in arm 2, the effective indices and arm lengths are

$$\begin{aligned}
n_{\text{eff},1}(\nu, \Delta T) &= n_{\text{eff}}(\nu) + \Delta n_{\text{eff}}(\nu, \Delta T) \\
n_{\text{eff},2}(\nu) &= n_{\text{eff}}(\nu) + \delta n(\nu) \\
L_1(\Delta T) &= L + \Delta L(\Delta T) \\
L_2 &= L + \delta L
\end{aligned} \quad (6)$$

The expressions for $n_{\text{eff}}(\nu)$, $\Delta n_{\text{eff}}(\nu, \Delta T)$, $\delta n(\nu)$ and $\Delta L(\Delta T)$ are presented in Suppl. Section 2 and contain high order terms in $\Delta T$ and/or $\Delta\nu$, with $\Delta\nu = \nu - \nu_0$.

The phase difference of the Si-FTS can be written in the form $2\pi \times frequency \times delay$, similar to eq.3 but with modified *frequency* and *delay* terms. Substituting eq.6 in eq.2 yields $\Delta\phi$ with contributions in $\Delta\nu^i\Delta T^j$ for $i$ from 0 to 4 and for $j$ from 0 to 5. The simplification of the resulting expression depends on the dispersion and thermo-optical properties of the waveguide as well as on the maximum temperature excursion and on the bandwidth of the light source. Using the properties of our 250 nm-by- 550 nm SOI waveguides, summarized in Suppl. Table 1, with temperature excursion of less than 100 K and optical bandwidth of a few tens of terahertz, we show that $\Delta\phi$ can be simplified to

$$\Delta\phi \approx \varphi(\nu) + 2\pi u\mathcal{T}. \quad (7)$$

The first term $\varphi(\nu)$ depends only on $\nu$ (see eq.S19) and therefore does not contribute to the kernel of the FT. It contributes to shifting and distorting the interferogram, but it has no influence on the retrieved PSD as will be shown. Thus, it is not discussed in detail here.

The second term, $2\pi u\mathcal{T}$ is similar to eq.3, but with modified optical frequency and time delay

$$\begin{aligned}
u &= \Delta\nu(1 + \xi_1) + \nu_0 \\
\mathcal{T} &= \tau + \gamma_2\tau^2.
\end{aligned} \quad (8)$$

$u$ represents the broadening of the original optical frequency $\nu$ around $\nu_0$ by a factor $1 + \xi_1$. In our case, $\xi_1$ is dominated by $\partial_{\nu,T} n$. $\mathcal{T}$ represents a correction of the original optical delay $\tau$ that includes the nonlinear contribution $\gamma_2\tau^2$, with $\gamma_2$ dominated by $\partial_{T^2} n$. $\tau$ in turn is related to $\Delta T$ by $\tau = \eta_1\Delta T$, with $\eta_1$ dominated by $\partial_T n$. The expressions for $\xi_1$, $\gamma_2$ and $\eta_1$ are presented in Suppl. Section 2.

Substituting eq.7 in eq.1 and changing the integration variable from $\nu$ to $u$, the interferogram and the PSD are finally related through a FT in the modified conjugate variables $u$ and $\mathcal{T}$, denoted by $\mathcal{F}[\;]$,

$$I(\mathcal{T}) = \frac{1}{1 + \xi_1}\mathcal{F}\left[e^{j\varphi(u)}T(u)\text{PSD}(u)\right] \quad (9)$$

Neglecting the constant term $(1+\xi_1)^{-1}$ multiplying the FT, the PSD is then retrieved from the absolute value of the IFT of the interferogram, normalized by the MZI transfer function $T(u)$,

$$\text{PSD}(u) = \frac{\left|\mathcal{F}^{-1}\left[I(\mathcal{T})\right]\right|}{T(u)}. \quad (10)$$

Finally, the frequency axis must be transformed back to the original optical frequency $\nu$,

$$\text{PSD}(u) \xrightarrow{\nu = \frac{u - \nu_0}{1 + \xi_1} + \nu_0} \text{PSD}(\nu). \quad (11)$$

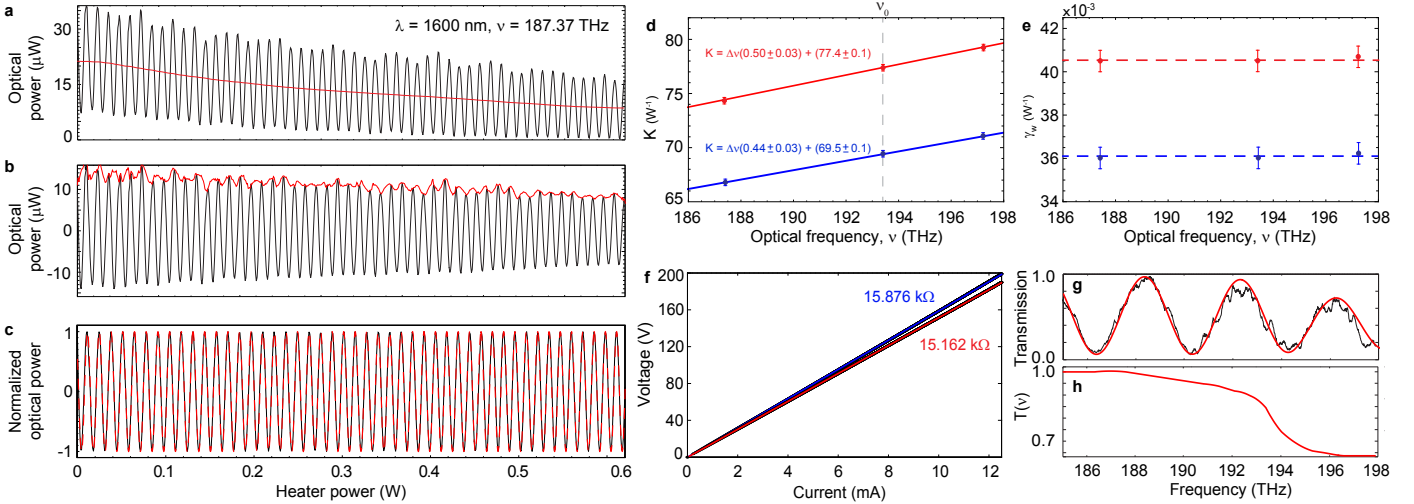

Figure 2: **Si-FTS calibration using a tunable laser source.** Measurements performed in the C-band. **a-f**: data in blue and red are related to heaters $H_1$ and $H_2$, respectively. **a.** As-measured interferogram at 183.37 THz (1600 nm) as a function of the heating power (W). To achieve robust non-linear fitting we subtract the mean power (red trace) and the envelope (red trace in **b**) to obtain the curve in **c**. **c.** The fitted curve (dashed-red trace) using eq.13 adjusts well the experimetal trace. **d.** Parameter $K(\nu)$ obtained from the non-linear fit, adjusted according to eq.14. **e.** Non-linear parameters $\gamma_{W,1} = (36.2 \pm 0.3) \cdot 10^{-3}$ $W^{-1}$ and $\gamma_{W,2} = (40.5 \pm 0.3) \cdot 10^{-3}$ $W^{-1}$ obtained from the non-linear fit. **f.** Current vs voltage (IV) response of both heaters and calculated electric resistance. The linearity of the IV curves confirm the thermo-optic origin of the non-linear behavior observed in the interferogram. **g.** Experimental (black trace) and calculated (red trace) transmission spectrum of the MZI at non-zero optical delay (17.2 ps). The calculated transmission is obtained using eq.15 in order to extract the MZI transfer function $T(\nu)$ shown in **h**.

## Spectrometer calibration with tunable laser

As in free-space, the Si-FTS must be calibrated to provide good absolute frequency accuracy. In addition, parameters $\xi_1$, $\gamma_2$ and $T(\nu)$ should also be ideally determined in a calibration step such that the transformations $u \to \nu$, $\tau \to \mathcal{T}$ and the renormalization of eq.10 can be properly performed in later use. A calibration process realized with a narrow linewidth tunable laser source allows to address all these requirements.

First, the calibration of the absolute optical frequency, $\xi_1$ and $\gamma_2$ is achieved measuring the interferogram of the laser source for various laser frequencies. Calibrating the absolute optical frequency reduces to determining $\kappa_\tau$ that connects the electric power dissipated in the heater with the resulting arm delay, $\tau = \kappa_\tau W$. The interferogram of a laser source at frequency $\nu$ is, according to eq.9,

$$I(\mathcal{T}) = A\cos\left(2\pi u \mathcal{T} + \varphi_\nu\right) \quad (12)$$

where $A$ and $\varphi_\nu$ are constant amplitude and phase. Using this relation and eqs.8, the interferogram can be written as a function of the electric power and the original laser frequency as

$$I(W,\nu) = A\cos\left[2\pi K(\nu)W(1+\gamma_W W) + \varphi_\nu\right] \quad (13)$$

with

$$\begin{aligned} K(\nu) &= \kappa_\tau (1+\xi_1)\,\Delta\nu + \kappa_\tau \nu_0 \\ \gamma_W &= \kappa_\tau \gamma_2. \end{aligned} \quad (14)$$

$K(\nu)$ and $\gamma_W$ can be determined for each heater ($H_1$ and $H_2$) curve-fitting the experimental interferograms using a cosine with non-linear argument. Following, the linear fit of $K(\nu)$ with $\nu_0 = 193.414$ THz allows to determine $\kappa_\tau$ for each heater and $\xi_1$. Finally, using $\kappa_\tau$ we obtain $\gamma_2 = \gamma_W/\kappa_\tau$.

The calibration results using the laser interferograms are presented in Fig.2a-e and the extracted parameters are summarized in Table 1. The interferogram for the laser frequency 187.37 THz (1600 nm) with heater $H_2$ actuated and its curve-fit are shown in Fig.2a-c as an example of the procedure realized for multiple frequencies and both heaters. Combining the interferogram fit results for each heater, $K(\nu)$ follows a linear dependence with frequency (Fig.2d) while $\gamma_W$ has a constant value (Fig.2e), in agreement with the eq.14.m The non-linear term $\gamma_2$ obtained separately for each heater has the same value within the experimental error, $(101 \pm 1) \cdot 10^{-3}$ $ps^{-1}$, while $\xi_1$ only slightly

| Table 1 | Calibration with a tunable laser |
|---|---|
| **Parameter** | **Value** |
| $\kappa_{\tau,1}$ ($10^{-3}$ ps.W$^{-1}$) | 359 ± 1 |
| $\kappa_{\tau,2}$ ($10^{-3}$ ps.W$^{-1}$) | 400 ± 1 |
| $\gamma_2$ ($10^{-3}$ ps$^{-1}$) | 101 ± 1 |
| $\xi_1$ ($10^{-2}$) | 23 ± 1 |

deviates for each heater, $0.22 \pm 0.02$ for $H_1$ and $0.24 \pm 0.01$ for $H_2$, yielding $0.23 \pm 0.02$.

The current versus voltage (IV) curves depicted in Fig.2f confirm the thermo-optic origin of the measured non-linearity and allow to understand the difference between $\kappa_{\tau,1}$ and $\kappa_{\tau,2}$. The IV plots show a fairly linear behavior for both heaters, with resistances 15,876 kΩ for $H_1$ and 15,162 kΩ for $H_2$, and indicates that any non-linearity originating from the heaters is small compared to the non-linearity intrinsic to the thermo-optic effect. Moreover, the difference in electric resistance causes a difference in heater efficiency $k_T$ that explains the observed discrepancy in the measured $\kappa_\tau$'s. The heater efficiency is determined such that the temperature change at the waveguide level $\Delta T$ is related to the dissipated electric power $W$ by $\Delta T = k_T W$, thus $\kappa_{\tau,i} = \eta_1 k_{T,i}$. The agreement in $\gamma_2$ and $\xi_1$ obtained independently for each heater indicates the two arms of the MZI are fairly similar, so that $\eta_1$ can be considered the same. Using $\eta_1 \approx 1.94 \cdot 10^{-2}$ ps/K obtained from simulations (see Suppl. Section 1), the heater efficiencies are estimated at $k_{T,1} \approx 18.5$ K/W and $k_{T,2} \approx 20.6$ K/W.

The interferometer transfer function $T(\nu)$ is obtained from the transmission spectrum of the MZI, as depicted in Fig.2g,h. The optical power at the output of the MZI is

$$I_{\text{out}}(\nu) = I_0(\nu) + T(\nu)\cos[\Delta\phi(\nu)]. \quad (15)$$

$T(\nu)$ is the envelope of the transmission oscillations and can be obtained by adjusting the experimental trace. It is recommended that the transmission spectrum be measured at a non-null phase difference $\Delta\phi$ so that $T(\nu)$ can be decoupled from the frequency-dependence of the average power $I_0(\nu)$. In our case, the experimental transmission (black trace) represents the transmission spectrum of the passive device when none of the heaters are actuated. The oscillations have a free-spectral range (FSR) of 3.96 THz around 193.414 THz (32.7 nm aroun 1550 nm) originated from slight differences between the two arms incorporated as $\delta L$ and $\delta n(\nu)$ as discussed in the previous section.

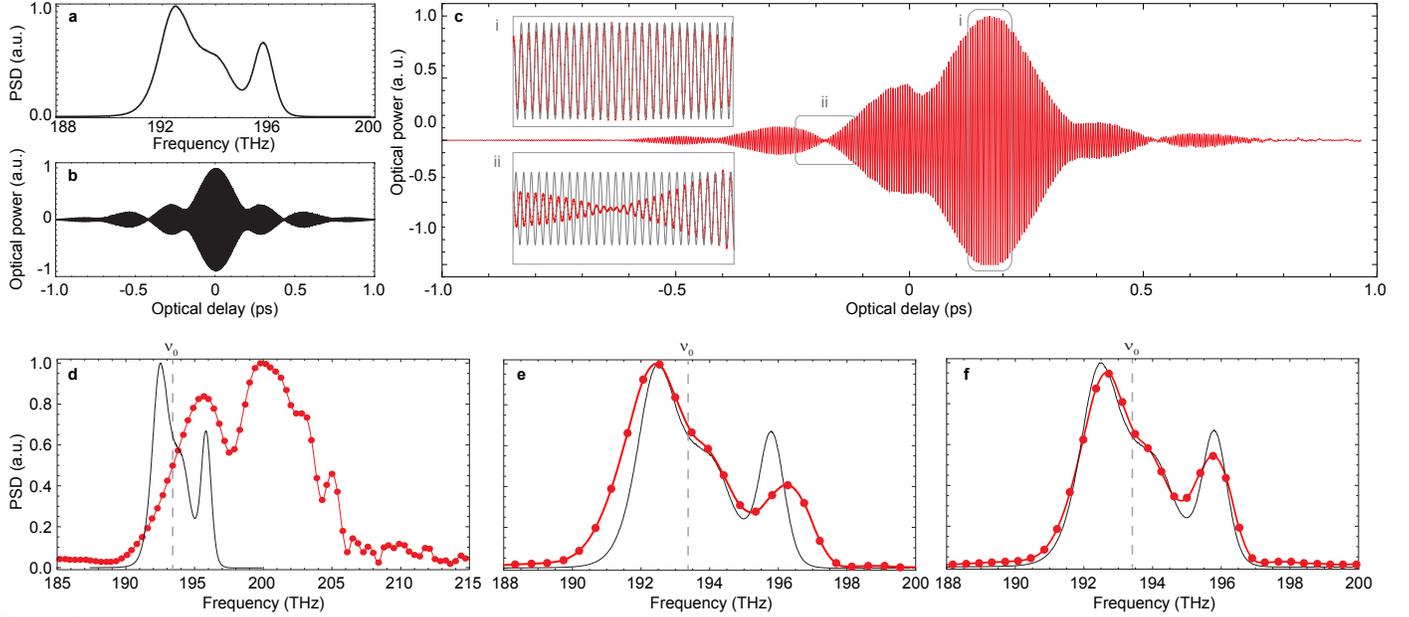

Figure 3: **Broadband spectrum recovery with the Si-FTS.** The parameters used to obtain these plots are summarized in Table 1. **a.** ASE of a C-band EDFA used as the light source. **b.** Theoretical interferogram of the ASE for an ideal (linear TOC, dispersionless, balanced) MZI. **c.** Experimental interferogram, shifted to 17.2 ps and distorted due to differences between the two arms of the MZI. The optical delay axis corresponds to $\mathcal{T}$. The insets show a zoom-in of the interferogram at different optical delays superposed to a cosine (grey traces) at the ASE mean frequency $\nu_s = 193.44$ THz, highlighting: **i** the oscillations at $\nu_s$ when the envelope varies slowly; **ii** phase changes when the envelope varies rapidly. **e-g.** Experimental (red) and reference (black) PSD at different conditions. The red points are the experimental data obtained from the IFT and the red line is a second-order interpolation curve. **d.** No correction, including thermo-optic non-linearity. **e.** Corrected thermo-optic non-linearity, but no dispersion correction or PSD re-normalization with $T(\nu)$. **f.** All effects properly accounted for.

## Spectrum recovery and discussion

The Si-FTS is validated by recovering the spectrum of the amplified spontaneous emission (ASE) of a C-band Erbium-Doped Fiber Amplifier (EDFA). The ASE provides a good test spectrum in the telecom band, suitable for testing with the available equipment in our lab. The reference ASE spectrum, measured with a tabletop optical spectrum analyzer (OSA), is shown in Fig.3a, and its theoretical ideal interferogram for a dispersionless, perfectly balanced Si-FTS is depicted in Fig.3b. The ASE presents two resolved peaks at 192.6 THz and 195.9 THz in addition to an unresolved peak at 194 THz, with total bandwidth around 7 THz (56 nm, 233.5 cm$^{-1}$).

The experimental interferogram is presented in Fig.3c. The delay axis corresponds to the transformed delay $\mathcal{T}$ obtained using the parameters from Table 1. Positive delay corresponds to heater $H_1$, while negative delay corresponds to $H_2$. It spans $\Delta\mathcal{T} = 2.13$ ps, corresponding to maximum dissipated powers around 2.6 W and 2.5 W in heaters $H_1$ and $H_2$ and maximum temperature excursions around 54 K and 46 K at the waveguide level.

A comparison between experimental and ideal interferograms shows some noticeable distinctions. The zero delay (maximum envelope amplitude) is shifted to around 17.2 ps and the interferogram envelope is asymmetric. These effects result from the non-zero contribution of $\varphi(\nu)$ in eq.9, which carries the contribution of the difference between arms, $\delta L$ and $\delta n(\nu)$ (see eq.S20). The zero-delay shift is caused by the first-order in $\nu$, while the envelope distortion is caused by high-order terms.

The difference between arms that cause these observable changes in the interferogram correspond to typical variations expected from the silicon photonics process, rather than strong differences due to a non-ideal fabrication process. Considering the difference in arm length $\delta L$ negligible, the zero delay is centered at

$$\mathcal{T}_0 \approx \frac{L}{c}\frac{\delta(n_{\text{eff}}|_{\nu_0}) + \nu_0\delta(\partial_\nu n)}{1+\xi_1}. \quad (16)$$

Using the known values for the other parameters, we estimate $n_{\text{eff}}|_{\nu_0} + \nu_0\delta(\partial_\nu n) \approx 2\cdot 10^{-3}$. This value agrees with the expected order of magnitude for effective index fluctuations due to chip-scale variations in the silicon device layer thickness[19,25]. The same is true for the differences in high-order dispersion terms.

The interferogram oscillations are highlighted in the insets of Fig.3c, where a cosine oscillation at the ASE mean frequency ($\nu_s = 193.44$ THz) is superposed (grey trace). In regions with a smooth envelope variation the interferogram oscillates at $\nu_s$ (inset i), while an additional phase is introduced as the envelope varies more abruptly (inset ii). The phase change due to envelope variations supports the interest of using a narrow laser source with almost flat envelope to calibrate the thermo-optic non-linearity, as the non-linear fit would be compromised by such phase change.

The PSD obtained from the experimental interferogram and the effects of thermo-optic non-linearity ($\gamma_2$), dispersion ($\xi_1$) and MZI transfer-function ($T(\nu)$) are presented in Fig.3d-f. The spectra were calculated using the mean value of the calibration parameters summarized in Table 1.

The PSD obtained directly form the as-measured interferogram – with the delay axis corresponding to $\tau$ and without performing any correction – is presented in Fig.3d. The TOC non-linearity distorts, broadens and shifts the PSD to higher frequencies as the interferogram oscillates faster with increasing delay change to higher frequencies.

After the optical delay axis of the interferogram is properly transformed to $\mathcal{T}$, the resulting PSD becomes very similar to the reference spectrum (Fig.3e). Both resolved peaks are clearly identified and the unresolved peak is also present around 194 THz. However, since the spectrum has not been re-scaled to the original frequency $\nu$, it is broadened by the factor $1+\xi_1$ around $\nu_0$. In addition, since it has not been re-normalized by $T(\nu)$, the high frequency peak appears attenuated relatively to the low frequency peak.

Failing to re-scale the frequency axis according to eq.11 introduces a frequency error $u - \nu = \xi_1 \Delta\nu$ that degrades the absolute frequency accuracy. In this demonstration, the effect is modest since the detuning of the peaks with respect to $\nu_0$ is small. The peak around 196 THz, for instance, is shifted by 0.5 THz, which is roughly its full-width at half maximum (FWHM). Nonetheless, the effect can be very significant for frequencies further from $\nu_0$.

The PSD corrected for the thermo-optic non-linearity, dispersion and the MZI transfer function reproduces satisfactorily well the reference spectrum (Fig.3f). This spectrum was obtained performing only the aforementioned corrections, with no additional data processing such as zero-filling or apodization of the interferogram. The spectral resolution of $\delta\nu = 0.38$ THz ($\delta\sigma = 12.7$ cm$^{-1}$, $\delta\lambda = 3.05$ nm) is comparable to other on-chip spectrometers aimed at broadband operation and it is sufficient for a large range of Raman and IR absorption spectroscopy applications [13,15].

It is worth noticing that the presence of TOC non-linearity and dispersion have an upside in the Si-FTS performance. Once $\gamma_2$ and $\xi_1$ are accounted and corrected for, they effectively increase the PSD resolution for a given dissipated power. In our case, the resolution of 0.38 THz, calculated as $\delta\nu_{\text{real}} = \Delta\mathcal{T}^{-1}(1+\xi_1)^{-1}$ corresponds to a 35% improvement with respect to a resolution of 0.515 THz calculated for an ideal Si-FTS as $\delta\nu_{\text{ideal}} = \Delta\tau^{-1}$ ($\gamma_2$ and $\xi_1$ are zero while all other parameters remain the same). The two resolutions are related by

$$\delta\nu_{\text{real}} = \frac{\delta\nu_{\text{ideal}}}{(1+\xi_1)(1+\gamma_2\Delta\tau/2)} \tag{17}$$

and the improvement can become very significant for large delays.

The ultimate performance of the on-chip Si-FTS is quite promising considering recent advancements in silicon photonics design and fabrication. Power couplers/splitters offering flat optical response over tens of terahertz [30,31] (a few hundreds of nanometers) may allow Si-FTS operating over large bandwidths. Moreover, extremely-long, low-loss silicon waveguides fabricated in tight footprints [27,32–34] may be employed and combined with high temperature excursions endured by CMOS-compatible silicon devices [35] to deliver very high spectral resolution while maintaining broadband operation. For instance, a Si-FTS with parameters similar to ours but with 64-cm long spiral arms [32] operated up to 350 K above room temperature should deliver extremely high-resolution around $\delta\nu = 1\cdot10^{-4}$ THz ($\delta\sigma = 3\cdot10^{-3}$ cm$^{-1}$, $\delta\lambda = 0.7$ pm). Finally, design changes may allow more efficient Si-FTS, such as using Michelson interferometers [36] instead of MZIs to double the optical path in a given footprint and introducing heat isolating structures to increase heater efficiency [37].

In addition to high performance, a valuable advantage of the Si-FTS compared to other on-chip spectrometer approaches is its robustness to fabrication variations. Although the interferogram is strongly affected by the difference in effective index between the arms of the MZI (Fig.3c), as previously discussed, the PSD remains unaffected (Fig.3f). This result is expected from our model since $\varphi(u)$ is canceled calculating the PSD by taking the absolute value of the IFT (eq.10).

In summary we demonstrated the realization of a Fourier Transform spectrometer with true time delay in silicon photonics. Considering the non-linearity of the thermo-optic effect as well as thermal expansion and dispersion we derived simple corrections that effectively account for these effects and allow to use well established FT techniques to obtain accurate spectral responses. We showed how the Si-FTS can be calibrated using a tunable laser source and we demonstrated the successful recovery of a broadband spectrum, resilient to fabrication variations. Our discussion proposes a simple approach to tackle the hurdles of doing FT spectrometry in silicon photonics and paves the way for robust, cost-effective and versatile FT-based portable spectrometers.

# Contributions

A.G. and Y.F. conceived and supervised the project. M.S and A.G. developed the mathematical model. N.F. co-supervised the project mostly with the mathematical model development. M.S. performed simulations, fabricated the samples and realized the experiments. All authors contributed to discussing the results and preparing the manuscript.

# Acknowledgments

This work was performed in part at the San Diego Nanotechnology Infrastructure (SDNI) of UCSD, a member of the National Nanotechnology Coordinated Infrastructure, which is supported by the National Science Foundation (Grant ECCS-1542148). M.S. and N.F. acknowledge financial support from the São Paulo Research Foundation (2014/04748-2,2015/20525-6). M.S. acknowledges fruitful discussions with colleagues from Fainman's group at UCSD and from the Device Research Laboratory at Unicamp.

# Methods

## Fabrication

The Si-FTS was fabricated in 15×15 cm$^2$ dies from 250-nm-thick SOI wafers with 3 mm of buried oxide. All the fabrication steps correspond to standard CMOS-compatible processes. The waveguides (550 nm wide) and input/output taper regions (150 nm wide at the tip) were patterned using electron-beam (e-beam) lithography with the negative tone resist HSQ. The separation between adjacent waveguides in the spiral sections is 5 $\mu$m center-to-center and the bending radius is 10 $\mu$m. After patterning, the waveguides are dry etched, the HSQ is striped with a dip in buffered oxide etch (BOE) solution and the waveguides are covered by 1.5 $\mu$m of PECVD-deposited silicon oxide cladding. After the oxide deposition, the heaters are patterned on top of each arm using PMMA and the metals are deposited via sputtering in a liftoff process. The heaters consist of a serpentine nichrome (NiCr) trail of 6 $\mu$m-wide 280 nm-thick sections separated by 4 $\mu$m border-to-border and totaling 17.33 mm in length. The NiCr heaters are terminated by 170×120 $\mu$m$^2$ titanium-gold (Ti:Au, 5 nm:300 nm) pads. A second PECVD silicon oxide layer (500 nm-thick) is deposited o top of the heaters and 100×100 $\mu$m$^2$ windows are opened on top of the Ti:Au pads to allow electric contact. After fabrication, the devices are diced and the input/output facets are polished to mitigate optical losses.

## Electrical and optical measurements

The calibration described in the main text was performed using a tunable laser source (Agilent 81600B) operating between 1460 nm and 1630 nm with optical power around 0 dBm. The EDFA broadband source (Amonics) delivered around 15 dBm of total optical power in the spectral range between 1528 nm and 1564 nm and was connected to an in-line fiber polarizer. Light was coupled into and out of the device using polarization maintaining (PM) cleaved fibers, oriented to excite the quasi-TE mode of the SOI waveguide. The output fiber was directly connected to an InGaAs powermeter. The coupling fibers were positioned using precise translation stages (Thorlabs NanoMax) driven by piezoelectric controllers. The fibers made physical contact with the sample, but were not permanently attached to it. The sample was mounted on top of a temperature stabilized station (20 C). The heaters were driven using a sourcemeter (Keithley 2400) allowing for simultaneous precise current drive (up to 2.1 A) and voltage monitoring (up to 210 V). The electric contact with the sample was done using micro-probes with 50×50 $\mu$m$^2$ tips connected to the heater pads. The total parasitic series resistance of the electric apparatus was less than 4 ohm. All the equipment was connected to a computer and controlled via Matlab routines. The interferograms were measured sweeping the electric current from zero to 2 A using a non-linear vector increasing with the square of the current such that the electric power vector would be linearly spaced. The interval between consecutive measurements was less than 0.5 seconds. Notice that a continuous current sweep was not employed only because the instruments used (Keithley 2400) did not support such continuous sweep.

# Supplementary Info: Fourier transform spectrometer on silicon with thermo-optic non-linearity and dispersion correction

Mario C. M. M. Souza[1,2], Andrew Grieco[1], Newton C. Frateschi[2], Yeshaiahu Fainman[1,*]

1. Department of Electrical and Computer Engineering, University of California, San Diego, 9500 Gilman Dr., La Jolla, California 92023, USA
2. "Gleb Wataghin" Physics Institute, University of Campinas, Campinas, SP 13083-970, Brazil

*email: fainman@ece.ucsd.edu

## 1 Dispersion, thermo-optic and thermal expansion parameters

In this section we present the properties of the silicon-on-insulator (SOI) waveguides used in our realization of the chip-scale silicon photonics Fourier Transform spectrometer (Si-FTS). Our design exploits the quasi-TE mode of a 250 nm-by 550-nm SOI waveguide. The effective index, its dispersion and its thermo-optic derivatives are summarized in Table 1. Here, we have used a contracted notation for partial derivatives, $\frac{\partial n_{\text{eff}}}{\partial x} \equiv \partial_x n$. These coefficients were obtained from finite difference element (FDE) simulations in the frequency range from 180 THz to 210 THz and for temperatures between 300 K and 400 K.

The simulations were performed using the FDE solver Lumerical MODE with modified bulk refractive index models for both the silicon core and the silica cladding.

The models include a Sellmeier dependence for both silicon[1] and silica[2]. The refractive index models presented in these references also include temperature dependence in the range from 20 K to 300 K, whereas we are interested in the temperature range from 300 K to 400 K. For the silica cladding, the thermo-optic coefficient (TOC) is practically constant around 300 K ($3 \cdot 10^{-5}$ K$^{-1}$), and it is safe to use the thermo-optic model provided in[2]. Silicon's TOC, on the other hand, varies significantly with temperature and the model provided in[1] does not deliver consistent results in the temperature range of interest. For this reason, we consider an index model for silicon that combines the dispersion from[1] with the TOC obtained from investigations comprising temperatures from 300 K to 400 K[3]. It is safe to introduce the thermo-optic behavior as independent additional terms in the refractive index model since the crossed-dependence $\frac{\partial^2 n_{\text{Si}}}{\partial \nu \partial T}$ is negligible in the frequency and temperature ranges of interest[4].

The thermo-optic behavior of silicon around 300 K is presented in Fig.1. In this model, the TOC has a second-order dependence with temperature (Fig.1a) and the contribution of the temperature dependent terms with respect to the zeroth order in depicted in Fig.1b. It shows that the first order has a non-negligible contribution reaching close to 14% of the zeroth order for a temperature change of 100 K, while the second order reaches a maximum of 1% in the same temperature range and can be neglected in practice.

In addition to the thermo-optic effect, thermal expansion changes the total length of the interferometer's arm and must be accounted for. The thermal expansion coefficient of silicon around 300 K is presented in Fig.2a and presents a strong dependence with temperature[5]. The temperature dependence is well described by a third-order polynomial with the coefficients presented as insets in 2a. The contribution of the temperature-dependent terms with respect to the zeroth order in depicted in Fig.2b. The first and second order terms contribute to 33% and 10% of the zeroth order value for temperature changes of 100 K. The third order reaches a maximum of 2% in the same temperature range and can be neglected. The coefficients used in our model (zeroth, first and second order) are also summarized in Table 1.

**Supplementary Table 1 |** Dispersion coefficients and thermo-optic coefficients for a 250-by-550 nm SOI strip waveguide at the telecom band ($\nu_0$ = 193.414 THz).

| Parameter | Value | Unit | Parameter | Value | Unit |
|---|---|---|---|---|---|
| $n_{\text{eff}}\vert_{\nu_0}$ | 2.62 | — | $\partial_{T^2} n_{\text{eff}}$ | $7.0 \times 10^{-7}$ | K$^{-2}$ |
| $\partial_{\nu} n_{\text{eff}}$ | $7.8 \times 10^{-3}$ | THz$^{-1}$ | $\partial_{\nu,T^2} n_{\text{eff}}$ | $-3.7 \times 10^{-10}$ | K$^{-2}$ THz$^{-1}$ |
| $\partial_{\nu^2} n_{\text{eff}}$ | $-9.0 \times 10^{-4}$ | THz$^{-2}$ | $\partial_{\nu^2,T^2} n_{\text{eff}}$ | $-3.7 \times 10^{-12}$ | K$^{-2}$ THz$^{-2}$ |
| $\partial_{\nu^3} n_{\text{eff}}$ | $1.6 \times 10^{-6}$ | THz$^{-3}$ | $\partial_{\nu^3,T^2} n_{\text{eff}}$ | $4.6 \times 10^{-15}$ | K$^{-2}$ THz$^{-3}$ |
| $\partial_T n_{\text{eff}}$ | $1.85 \times 10^{-4}$ | K$^{-1}$ | $\alpha_1$ | $2.5 \times 10^{-6}$ | K$^{-1}$ |
| $\partial_{\nu,T} n_{\text{eff}}$ | $2.1 \times 10^{-7}$ | K$^{-1}$ THz$^{-1}$ | $\alpha_2$ | $8.5 \times 10^{-9}$ | K$^{-2}$ |
| $\partial_{\nu^2,T} n_{\text{eff}}$ | $-2.0 \times 10^{-9}$ | K$^{-1}$ THz$^{-2}$ | $\alpha_3$ | $-2.3 \times 10^{-11}$ | K$^{-3}$ |
| $\partial_{\nu^3,T} n_{\text{eff}}$ | $1.8 \times 10^{-10}$ | K$^{-1}$ THz$^{-3}$ | | | |

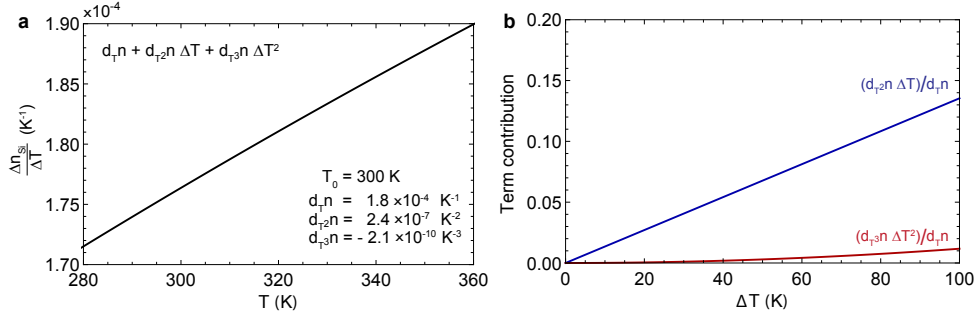

Figure 1: **Thermo-optic coefficient (TOC) of silicon**. **a.** Around 300 K, the TOC has a second order dependence with temperature with coefficients shown in the figure. **b.** Contribution of first and second order terms with respect to zeroth order. For a temperature change of 100 K, the first order contributes around 14%. The second order contribution is around 1% and can be neglected. The model is based on [3].

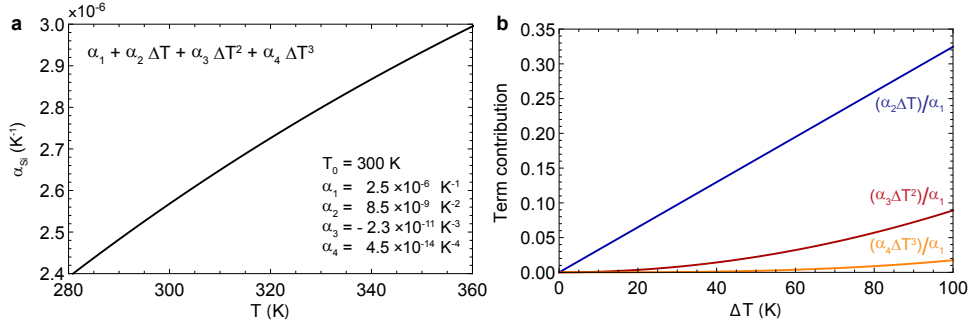

Figure 2: **Thermal expansion coefficient of silicon**. **a.** Around 300 K it has a strong dependence with temperature, modeled using a third order polynomial with the coefficients shown in the figure. **b.** Contribution of first, second and third order terms with respect to zeroth order. For a temperature change of 100 K, the first and second order contribute around 33% and 10%. The third order contribution is around 2% and can be neglected. The model is based on [5].

## 2 Interferogram and power spectral density for the Si-FTS

In this section we derive a Fourier Transform (FT) relation between the varying optical power at the output of a Mach-Zehnder interferometer (MZI), $I_{\text{out}}$, and the power spectral density of a broadband source, PSD($\nu$). In the ideal FTS the kernel of the FT is a complex exponential with argument $2\pi\nu\tau$, where $\nu$ is the optical frequency and $\tau$ is the time delay between the two arms of the MZI. We show that for the Si-FTS the kernel can still be presented with an argument of the form $2\pi \times frequency \times delay$, provided that *frequency* and *delay* are modified to account for dispersion, thermo-optic non-linearity and thermal expansion. In this general treatment, fabrication imperfections are also considered.

The simplifications carried in this section depend on various waveguide parameters and therefore are valid for the quasi-TE mode of a 250 nm-by-550 nm SOI waveguide and similar designs. They also depend on the temperature range achieved with the heaters and the bandwidth of the incoming light, which are considered close to our experimental conditions, namely, temperature excursion lower than 100 K and bandwidth around 10 THz. Nonetheless, our discussion provides a general procedure that is readily applicable to other device designs and operation conditions and allows to identify terms that might need to be incorporated as corrections.

The real electric field of an incoming broadband source can be written as

$$E(t) = \int_{-\infty}^{+\infty} S(\nu)e^{j2\pi\nu t}d\nu \tag{S1}$$

where $S(\nu)$ is the Fourier Transform of $E(t)$,

$$S(\nu) = \int_{-\infty}^{+\infty} E(t)e^{-j2\pi\nu t}dt \tag{S2}$$

Since $E(t)$ is real, it follows that

$$S(\nu) = S^*(-\nu). \tag{S3}$$

Neglecting proportionality constants, the power spectral density can be defined as [6]

$$\text{PSD}(\nu) \equiv |S(\nu)|^2. \tag{S4}$$

The MZI is composed of input and output couplers with frequency-dependent field coupling coefficients $(t_1(\nu), k_1(\nu))$ and $(t_2(\nu), k_2(\nu))$, respectively, and two arms with attenuation factors $R_1(\nu)$ and $R_2(\nu)$. Thus, after splitting into arms 1 and 2 and recombining at the output waveguide, the electric field that propagated through each arm is

$$E_1(t) = \int_{-\infty}^{+\infty} T_1(\nu)S(\nu)e^{j2\pi\nu t}e^{j\phi_1(\nu)}d\nu \tag{S5}$$

$$E_2(t) = \int_{-\infty}^{+\infty} T_2(\nu)S(\nu)e^{j2\pi\nu t}e^{j\phi_2(\nu)}d\nu \tag{S6}$$

where $T_1(\nu) = t_1(\nu)t_2(\nu)R_1(\nu)$ and $T_2(\nu) = k_1(\nu)k_2(\nu)R_2(\nu)$ are real transmission factors and $\phi_i(\nu)$ is the real phase accumulated through propagation in arm $i$, given by

$$\phi_i(\nu) = \beta_i(\nu)L_i = \frac{2\pi\nu}{c} n_{\text{eff,i}}(\nu)L_i \tag{S7}$$

where $\beta_i$, $L_i$ and $n_{\text{eff,i}}$ are the propagation constant, length and effective index of arm $i$ and $c$ is the speed of light in vacuum. The real nature of the output fields requires that

$$\begin{aligned} T_i(\nu) &= T_i(-\nu) \\ \phi_i(\nu) &= -\phi_i(-\nu) \longrightarrow n_{\text{eff,i}}(\nu) = n_{\text{eff,i}}(-\nu). \end{aligned} \tag{S8}$$

The output power is then obtained time averaging the squared output electric field through many field oscillations,

$$\begin{aligned} I_{\text{out}} &\propto \langle E_{out}^2(t) \rangle \\ &\propto \langle [E_1(t) + E_2(t)]^2 \rangle \\ &\propto \langle E_1^2(t) \rangle + \langle E_2^2(t) \rangle + 2\langle E_1(t)E_2(t) \rangle. \end{aligned} \tag{S9}$$

The first two terms in the last row of eq.S9 contribute to the mean output power, while the last term contains the interference term of interest. In the following we will focus on manipulating such term, which after some manipulation can be written as

$$I = 2\langle E_1(t)E_2(t)\rangle \propto \int_{-\infty}^{+\infty} T(\nu)\text{PSD}(\nu)e^{j\Delta\phi(\nu)}d\nu \tag{S10}$$

with $T(\nu) = T_1(\nu)T_2(\nu)$ and

$$\Delta\phi = \frac{2\pi\nu}{c}\left[n_{\text{eff},1}(\nu)L_1 - n_{\text{eff},2}(\nu)L_2\right]. \tag{S11}$$

Dropping the proportionality sign and taking $I$ equal to the right hand side (RHS) of eq.S10 gives the relation between the oscillatory output power from the MZI – the interferogram – and the PSD in the general case.

The phase difference $\Delta\phi$ is determined by the difference between the optical path ($n_{\text{eff}}L$) of the two arms, as indicated by eq.S11. By design the interferometer is balanced ($\Delta\phi = 0$) when no heating is applied, with both arms having the same effective index $n_{\text{eff}}(\nu)$ and arm length $L$. Typical silicon waveguides present strong dispersion, which are accounted for in a series expansion of the effective index around the central optical frequency $\nu_0$ (193.414 THz in our case) up to third order,

$$n_{\text{eff}}(\nu) = \left.n_{\text{eff}}\right|_{\nu_0} + \partial_\nu n\,\Delta\nu + \frac{1}{2}\partial_{\nu^2}n\,\Delta\nu^2 + \frac{1}{6}\partial_{\nu^3}n\,\Delta\nu^3, \tag{S12}$$

using the contracted notation for partial derivatives, $\frac{\partial n_{\text{eff}}}{\partial x} \equiv \partial_x n$ and with $\Delta\nu = \nu - \nu_0$.

In practice, chip-scale variations in the device layer thickness and fabrication imperfections will cause differences between the two arms. These differences, considered small, are introduced as "$\delta$" terms. For instance, the difference in effective index at $\nu_0$ and its first derivative are denoted by $\delta(n_{\text{eff}}|_{\nu_0})$ and $\delta(\partial_\nu n)$ respectively, while the difference introduced in arm length is $\delta L$.

When the heaters are actuated, the temperature change $\Delta T$ modifies the effective index through the thermo-optic effect and the length of the silicon waveguide arm through thermal expansion. For the Si-FTS, it is important to consider the high order temperature dependence of both effects, as well as the dispersion of the thermo-optic effect. As discussed in the previous section, for temperatures rising up to 100 K above room temperature (i.e. up to ∼400 K), it suffices to consider the first and the second order temperature-dependence of the TOC and of the thermal expansion coefficient, respectively. These lead to a second order and third-order temperature dependence of the effective index change $\Delta n_{\text{eff}}$ and of the arm length change $\Delta L$,

$$\begin{aligned} \Delta n_{\text{eff}}(\nu, \Delta T) =\ & \left(\partial_T n + \partial_{\nu,T}n\,\Delta\nu + \frac{1}{2}\partial_{\nu^2,T}n\,\Delta\nu^2 + \frac{1}{6}\partial_{\nu^3,T}n\,\Delta\nu^3\right)\Delta T\ + \\ & \frac{1}{2}\left(\partial_{T^2}n + \partial_{\nu,T^2}n\,\Delta\nu + \frac{1}{2}\partial_{\nu^2,T^2}n\,\Delta\nu^2 + \frac{1}{6}\partial_{\nu^3,T}n\,\Delta\nu^3\right)\Delta T^2 \end{aligned} \tag{S13}$$

$$\Delta L(\Delta T) = L\left(\alpha_1\Delta T + \alpha_2\Delta T^2 + \alpha_3\Delta T^3\right). \tag{S14}$$

Modifications of the effective index due to thermal expansion of the waveguide in its cross-section is negligible compared to the other effects and are not included.

With no loss of generality, we introduce temperature effects in arm 1 while we include the differences of effective index and arm length in arm 2, such that

$$n_{\text{eff},1}(\nu, \Delta T) = n_{\text{eff}}(\nu) + \Delta n_{\text{eff}}(\nu, \Delta T) \qquad n_{\text{eff},2}(\nu) = n_{\text{eff}}(\nu) + \delta n(\nu) \tag{S15}$$

$$L_1(\Delta T) = L + \Delta L(\Delta T) \qquad L_2 = L + \delta L \tag{S16}$$

with

$$\delta n(\nu) = \delta(n_{\text{eff}}|_{\nu_0}) + \delta(\partial_\nu n)\,\Delta\nu + \frac{1}{2}\delta(\partial_{\nu^2}n)\,\Delta\nu^2 + \frac{1}{6}\delta(\partial_{\nu^3}n)\,\Delta\nu^3 \tag{S17}$$

The phase difference $\Delta\phi$ can be expressed in an efficient way to sort and evaluate the contribution of the various orders of frequency detuning $\Delta\nu$ and time delay $\tau$ introduced by the temperature difference $\Delta T$. Substituting eqs.S12-S17 into eq.S11 and manipulating the resulting expression we write

$$\Delta\phi(\nu, \tau) = \varphi(\nu) + 2\pi\sum_{i=1}^{5}\frac{\eta_i}{\eta_1^i}\tau^i[\nu_0 + \Delta\nu(1+\xi_i)]\left[1 + \left(\frac{\chi_i + \xi_i}{1+\xi_i}\right)\frac{\Delta\nu}{\nu_0} + \left(\frac{\chi_i + \zeta_i}{1+\xi_i}\right)\left(\frac{\Delta\nu}{\nu_0}\right)^2 + \left(\frac{\zeta_i}{1+\xi_i}\right)\left(\frac{\Delta\nu}{\nu_0}\right)^3\right]. \tag{S18}$$

The first term, $\varphi$, gathers frequency-dependent terms that do not depend on $\tau$. It is given by

$$\varphi(\nu) = -2\pi\ t_0 \left[\nu_0\sigma_0 + (\nu_0\sigma_1 + \sigma_0)\ \Delta\nu + (\nu_0\sigma_2 + \sigma_1)\ \Delta\nu^2 + (\nu_0\sigma_3 + \sigma_2)\ \Delta\nu^3 + \sigma_3\ \Delta\nu^4\right] \quad (S19)$$

with $t_0 \equiv \frac{L}{c}$ and

$$\sigma_0 = \delta(n_{\text{eff}}|_{\nu_0}) + \frac{\delta L}{L} n_{\text{eff}}|_{\nu_0} \quad \text{and} \quad \sigma_i = \delta(\partial_{\nu^i} n) + \frac{\delta L}{L}\partial_{\nu^i} n \quad (i=1,2,3) \quad (S20)$$

Assuming $\sigma_{-1} = \sigma_4 = 0$, eq.S19 can be recast in a compact form as

$$\varphi(\nu) = \sum_{i=0}^{4} t_0(\sigma_{i-1} + \nu_0\sigma_i)\ \Delta\nu^i. \quad (S21)$$

With no dependence in $\tau$, $\varphi$ does not contribute to the kernel of the transformation between time and frequency, on the other hand, it is the only term with contribution from $\delta$-terms.

For now, we focus on the summation containing the *frequency*$\times$*delay* terms of interest. In this form, the summation gathers the remaining terms by increasing order of $\tau$. The terms $\tau^i$ are weighted by $\eta_i/\eta_1^i$, and the time delay itself is related to the temperature difference by $\tau = \eta_1 \Delta T$. Each $\eta_i$ gathers terms in $T^{-i}$, multiplied by the characteristic time $\tau_0 \equiv \frac{L}{c}$,

$$\eta_1 = t_0(n_{\text{eff}}|_{\nu_0}\ \alpha_1 + \partial_T n),$$
$$\eta_2 = t_0(n_{\text{eff}}|_{\nu_0}\ \alpha_2 + \partial_T n\ \alpha_1 + \frac{1}{2}\partial_{T^2} n)$$
$$\eta_3 = t_0(n_{\text{eff}}|_{\nu_0}\ \alpha_3 + \partial_T n\ \alpha_2 + \frac{1}{2}\partial_{T^2} n\ \alpha_1) \quad (S22)$$
$$\eta_4 = t_0(\partial_T n\ \alpha_3 + \frac{1}{2}\partial_{T^2} n\ \alpha_2)$$
$$\eta_5 = t_0(\frac{1}{2}\partial_{T^2} n\ \alpha_3).$$

Assuming $\partial_{T^0} n \equiv n_{\text{eff}}|_{\nu_0}$ and $\alpha_0 \equiv 1$, these terms can be written in a general form as

$$\eta_k = t_0 \sum_{i=0}^{k} \frac{1}{i!}\partial_{T^i} n\ \alpha_{k-i}. \quad (S23)$$

The contribution of high order terms compared to the first order is depicted in Fig.3 as a function of the time delay. These plots were obtained using the parameters summarized in Supplementary Table 1. The second order (blue trace) contributes significantly and therefore cannot be neglected even for small delays. The third order term can be neglected for small delays, but it becomes increasingly important for large delays required for high spectral resolution. It reaches 1% of the first order contribution at 3.5 ps, and $\sim 10\%$ at 10 ps (Fig.3a). Higher order terms are very small and can be neglected even for large delays. In our experiment, we achieve a maximum time delay around 1 ps, in which regime it suffices to keep the first and second order contributions (Fig.3b).

Multiplying each order in $\tau$ in eq.S18 we have terms of increasing order in $\Delta\nu$. The first brackets $[\nu_0 + \Delta\nu(1+\xi_i)]$ represent zeroth and first order contributions, while the second brackets assemble higher orders terms. The parameters $\xi_i$, $\chi_i$ and $\zeta_i$ are adimensional and represent, respectively, the first, second and third order dispersion terms in $T^{-i}$:

$$\begin{aligned}
\xi_1 &= \frac{t_0\nu_0}{\eta_1}(\partial_\nu n\ \alpha_1 + \partial_{\nu,T} n) & \chi_1 &= \frac{1}{2}\frac{t_0\nu_0^2}{\eta_1}(\partial_{\nu^2} n\ \alpha_1 + \partial_{\nu^2,T} n) & \zeta_1 &= \frac{1}{6}\frac{t_0\nu_0^3}{\eta_1}(\partial_{\nu^3} n\ \alpha_1 + \partial_{\nu^3,T} n) \\
\xi_2 &= \frac{t_0\nu_0}{\eta_2}(\partial_\nu n\ \alpha_2 + \partial_{\nu,T} n\ \alpha_1 + \frac{1}{2}\partial_{\nu,T^2} n) & \chi_2 &= \frac{1}{2}\frac{t_0\nu_0^2}{\eta_2}(\partial_{\nu^2} n\ \alpha_2 + \partial_{\nu^2,T} n\ \alpha_1 + \frac{1}{2}\partial_{\nu^2,T^2} n) & \zeta_2 &= \frac{1}{6}\frac{t_0\nu_0^3}{\eta_2}(\partial_{\nu^3} n\ \alpha_2 + \partial_{\nu^3,T} n\ \alpha_1 + \frac{1}{2}\partial_{\nu^3,T^2} n) \\
\xi_3 &= \frac{t_0\nu_0}{\eta_3}(\partial_\nu n\ \alpha_3 + \partial_{\nu,T} n\ \alpha_2 + \frac{1}{2}\partial_{\nu,T^2} n\ \alpha_1) & \chi_3 &= \frac{1}{2}\frac{t_0\nu_0^2}{\eta_3}(\partial_{\nu^2} n\ \alpha_3 + \partial_{\nu^2,T} n\ \alpha_2 + \frac{1}{2}\partial_{\nu^2,T^2} n\ \alpha_1) & \zeta_3 &= \frac{1}{6}\frac{t_0\nu_0^3}{\eta_3}(\partial_{\nu^3} n\ \alpha_3 + \partial_{\nu^3,T} n\ \alpha_2 + \frac{1}{2}\partial_{\nu^3,T^2} n\ \alpha_1) \\
\xi_4 &= \frac{t_0\nu_0}{\eta_4}(\partial_{\nu,T} n\ \alpha_3 + \frac{1}{2}\partial_{\nu,T^2} n\ \alpha_2) & \chi_4 &= \frac{1}{2}\frac{t_0\nu_0^2}{\eta_4}(\partial_{\nu^2,T} n\ \alpha_3 + \frac{1}{2}\partial_{\nu^2,T^2} n\ \alpha_2) & \zeta_4 &= \frac{1}{6}\frac{t_0\nu_0^3}{\eta_4}(\partial_{\nu^3,T} n\ \alpha_3 + \frac{1}{2}\partial_{\nu^3,T^2} n\ \alpha_2) \\
\xi_5 &= \frac{t_0\nu_0}{\eta_5}(\frac{1}{2}\partial_{\nu,T^2} n\ \alpha_3) & \chi_5 &= \frac{1}{2}\frac{t_0\nu_0^2}{\eta_5}(\frac{1}{2}\partial_{\nu^2,T^2} n\ \alpha_3) & \zeta_5 &= \frac{1}{6}\frac{t_0\nu_0^3}{\eta_5}(\frac{1}{2}\partial_{\nu^3,T^2} n\ \alpha_3).
\end{aligned} \quad (S24)$$

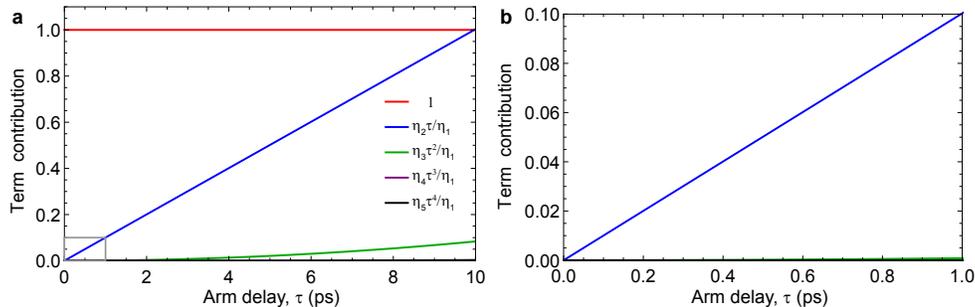

Figure 3: Contribution of terms is $\eta_i\tau^i/\eta_1^i$ with respect to the first order $\eta_1\tau$ as a function of the time delay in the range **a.** between 0 and 10 ps and **b.** between 0 and 1 ps – range of our experiments.

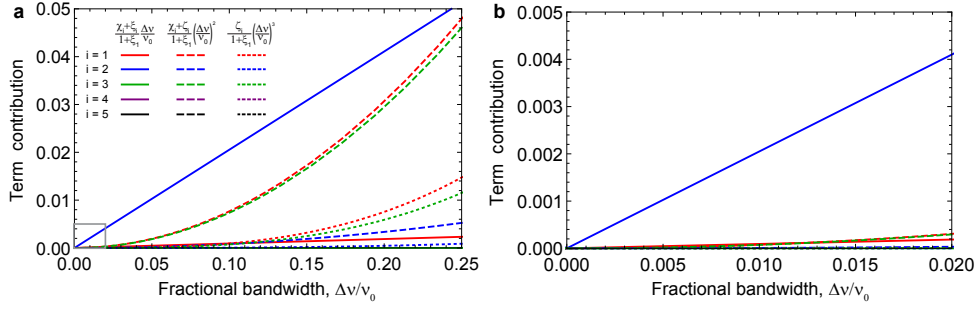

Figure 4: Contribution of each term in the second brackets of eq.S18 as a function of the fractional bandwidth in the range **a.** between 0 and 0.25 –large fractional bandwidth – and **b.** between 0 and 0.02 – range of our experiments.

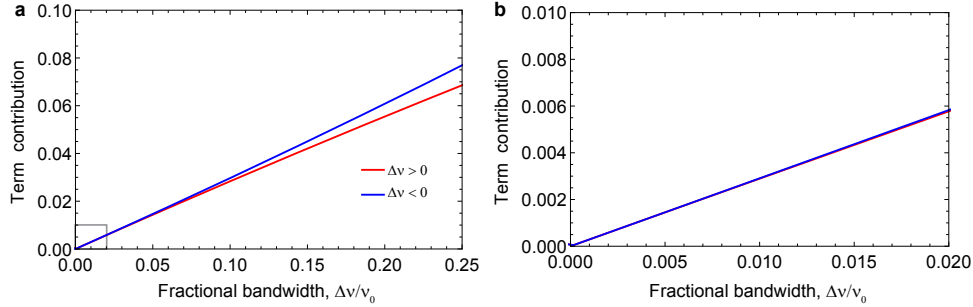

Figure 5: Contribution of $F_\xi$ in eq.S27 as a function of the fractional bandwidth in the range **a.** between 0 and 0.25 –large fractional bandwidth – and **b.** between 0 and 0.02 – range of our experiments.

They can be presented in a compact form analogous to eq.S23 assuming $\partial_{\nu^i,T^0} n \equiv \partial_{\nu^i} n$, $\alpha_0 \equiv 1$ and $\alpha_4 = \alpha_5 \equiv 0$:

$$\xi_k = \frac{\nu_0}{\eta_k}\frac{\partial \eta_k}{\partial \nu} = \frac{t_0 \nu_0}{\eta_k}\sum_{i=0}^{k}\frac{1}{i!}\partial_{\nu,T^i} n\ \alpha_{k-i}$$

$$\chi_k = \frac{\nu_0^2}{\eta_k}\frac{\partial^2 \eta_k}{\partial \nu^2} = \frac{1}{2}\frac{t_0 \nu_0^2}{\eta_k}\sum_{i=0}^{k}\frac{1}{i!}\partial_{\nu^2,T^i} n\ \alpha_{k-i} \quad \text{(S25)}$$

$$\zeta_k = \frac{\nu_0^3}{\eta_k}\frac{\partial^3 \eta_k}{\partial \nu^3} = \frac{1}{6}\frac{t_0 \nu_0^3}{\eta_k}\sum_{i=0}^{k}\frac{1}{i!}\partial_{\nu^3,T^i} n\ \alpha_{k-i}.$$

The contribution of the various terms in the second brackets is depicted in Fig.4 as a function of the fractional bandwidth $\Delta\nu/\nu_0$. For $\nu_0 = 193.414$ THz, a fractional bandwidth of 0.25, the upper limit of Fig.4a, corresponds to a frequency range from 145 THz to 240 THz (wavelength range from 1160 nm to 1930 nm), encompassing the full range of operation of expected Si-FTS designs operating around this $\nu_0$. Even in such large frequency range the contribution of high order terms is overall small, with individual terms reaching a maximum of 5% of the linear term (corresponding to 1). From Fig.4a we identify the terms in order $\tau^2\Delta\nu^2$ (blue trace), $\tau\Delta\nu^3$ (red-dashed trace) and $\tau^3\Delta\nu^3$ (green-dashed trace) as those with increasingly significant contribution for large fractional bandwidth, while the other terms are negligible. In our experiments we used a broadband source with total bandwidth around 7 THz centered at 193.44 THz, with a maximum fractional bandwidth around 0.2. In this case, depicted in Fig.4b, all the contributions are smaller than 0.5% and considered negligible.

Based on the these results (Fig.3,4), we keep only the first and second order terms in $\tau$ and we neglect the high orders in $\Delta\nu$ altogether, leaving only the unity in the second brackets of eq.S18.

The phase difference then simplifies to

$$\Delta\phi(\nu,\tau) = \varphi(\nu) + 2\pi\{\tau[\nu_0 + \Delta\nu(1+\xi_1)] + \frac{\eta_2}{\eta_1^2}\tau^2[\nu_0 + \Delta\nu(1+\xi_2)]\}. \quad \text{(S26)}$$

Defining $\gamma_2 \equiv \eta_2/\eta_1^2$ and rearranging the expression,

$$\Delta\phi(\nu,\tau) = \varphi(\nu) + 2\pi\left\{\left[\tau + \gamma_2\tau^2(1+F_\xi)\right][\nu_0 + \Delta\nu(1+\xi_1)]\right\}, \quad \text{(S27)}$$

with

$$F_\xi = \frac{(\xi_2 - \xi_1)\frac{\Delta\nu}{\nu_0}}{1+\xi_1\frac{\Delta\nu}{\nu_0}}. \quad \text{(S28)}$$

Once again we have an adimensional parameter $F_\xi$ to be compared to 1 as a function of the fractional bandwidth (Fig.5). $F_\xi$ is slightly different depending on the sign of $\Delta\nu$ as indicated in Fig.5a and reaches a maximum around 8% at $\frac{\Delta\nu}{\nu_0} = 0.25$, at which point corrections related to high order terms in $\tau\Delta\nu$ might also be considered as previously discussed. Close to our experimental conditions, however, the term is smaller than 1% (Fig.5b) and can be neglected.

Finally, the phase difference assumes the simple form

$$\Delta\phi(\nu,\tau) = \varphi(\nu) + 2\pi u \mathcal{T}, \tag{S29}$$

with

$$\begin{aligned}\mathcal{T} &\equiv \tau + \gamma_2 \tau^2 \\ u &\equiv \nu_0 + \Delta\nu(1+\xi_1)\end{aligned} \tag{S30}$$

as modified time delay and optical frequency. $\mathcal{T}$ linearizes the delay metrics, while $u$ stretches the original frequency $\nu$ around $\nu_0$ by a factor $(1+\xi_1)$.

Substituting eqs.S29,S30 in eq.S10 and performing the change of variables from $\nu$ to $u$ in the integral,

$$\begin{aligned}I(\mathcal{T}) &= \frac{1}{1+\xi_1} \int_{-\infty}^{+\infty} T(u) \text{PSD}(u) e^{j\varphi(u)} e^{j2\pi u \mathcal{T}} du \\ &= \frac{1}{1+\xi_1} \mathcal{F}\left[T(u)\text{PSD}(u)e^{j\varphi(u)}\right]\end{aligned} \tag{S31}$$

where $\mathcal{F}[\ ]$ denotes the Fourier Transform. Neglecting the constant term $(1+\xi_1)^{-1}$ multiplying the FT, the PSD is then retrieved from the absolute value of the inverse-FT of the interferogram, normalized by the MZI transfer function $T(u)$,

$$\text{PSD}(u) = \frac{\left|\mathcal{F}^{-1}\left[I(\mathcal{T})\right]\right|}{T(u)}. \tag{S32}$$

Finally, the frequency axis must be transformed back to the original optical frequency $\nu$,

$$\text{PSD}(u) \xrightarrow{\nu = \frac{u-\nu_0}{1+\xi_1} + \nu_0} \text{PSD}(\nu). \tag{S33}$$